\documentstyle[twoside]{article}
\newcount\Mac  \Mac=0  
\newcommand{\ifMac}[2]{\ifnum\Mac=1 #1 \else #2 \fi}
\newcommand{\riga}[1]{\noalign{\hbox{\parbox{\textwidth}{#1}}}\nonumber}
\newcommand{\bAk}[3]{\langle #1|#2|#3\rangle}
\newcommand{\One}{\hbox{1\kern-.24em I}}

\newcommand{\M}{{\cal M}}
\newcommand{\GeV}{\,{\rm GeV}}

\newcommand{\NP}{Nucl. Phys.}
\newcommand{\PRL}{Phys. Rev. Lett.}
\newcommand{\PL}{Phys. Lett.}
\newcommand{\PR}{Phys. Rev.}
\newcommand{\pL}{{\cal P}_{\rm L}} 
\newcommand{\pR}{{\cal P}_{\rm R}}

\newcommand{\eq}[1]{~{\rm (\ref{eq:#1})}}

\newcommand{\lnEps}{\ln\frac{\mub^2}{m_t^2}}
\newcommand{\epsIR}{\varepsilon_{\rm ir}}
\newcommand{\epsUV}{\varepsilon_{\rm uv}}
\newcommand{\eps}{\varepsilon}
\newcommand{\mub}{\bar{\mu}}
\newcommand\Ord{{\cal O}}
\newcommand\Op[1]{{\cal O}_{#1}}

\newcommand{\ds}{\partial\!\!\!\raisebox{2pt}[0pt][0pt]{$\scriptstyle/$}\,}

\newcommand{\Li}{\hbox{Li}_2}

\def\Red{}
\def\Black{}
\def\Blue{}

\newcommand{\lascia}[1]{}
\makeatletter
\def\art{\@ifnextchar[{\eart}{\oart}}
\def\eart[#1]#2#3#4#5#6{{\rm #2}, {\em #3 \bf #4} {\rm (#6) #5}}
\def\hepart[#1]#2{{\rm #2, \em#1}}
\newcommand{\oart}[5]{{\rm #1}, {\em #2 \bf #3} {\rm (#5) #4}}
\newcommand{\y}{{\rm and} }

\newcounter{alphaequation}[equation]
\def\thealphaequation{\theequation\hbox to
0.6em{\hfil\alph{alphaequation}\hfil}}
\def\eqnsystem#1{
\def\@eqnnum{{\rm (\thealphaequation)}}
\def\@@eqncr{\let\@tempa\relax \ifcase\@eqcnt \def\@tempa{& & &} \or
  \def\@tempa{& &}\or \def\@tempa{&}\fi\@tempa
  \if@eqnsw\@eqnnum\refstepcounter{alphaequation}\fi
\global\@eqnswtrue\global\@eqcnt=0\cr}
\refstepcounter{equation} \let\@currentlabel\theequation \def\@tempb{#1}
\ifx\@tempb\empty\else\label{#1}\fi
\refstepcounter{alphaequation}
\let\@currentlabel\thealphaequation
\global\@eqnswtrue\global\@eqcnt=0 \tabskip\@centering\let\\=\@eqncr
$$\halign to \displaywidth\bgroup \@eqnsel\hskip\@centering
$\displaystyle\tabskip\z@{##}$&\global\@eqcnt\@ne
\hskip2\arraycolsep\hfil${##}$\hfil& \global\@eqcnt\tw@\hskip2\arraycolsep
$\displaystyle\tabskip\z@{##}$\hfil
\tabskip\@centering&\llap{##}\tabskip\z@\cr}

\def\endeqnsystem{\@@eqncr\egroup$$\global\@ignoretrue} \makeatother

\def\Ord{{\cal O}}

\def\SU{{\rm SU}}

\def\circa#1{\,\raise.3ex\hbox{$#1$\kern-.75em\lower1ex\hbox{$\sim$}}\,}

\begin{document}
\begin{quote}
{\em 12 Oct 1997}\hfill {\bf IFUP--TH 48/97}\\
\phantom{.} \hfill{\bf hep-ph/9710312}
\end{quote}
\bigskip
\centerline{\huge\bf\Red Two-loop QCD corrections to}
\centerline{\huge\bf\Red charged-Higgs-mediated $b\to s\gamma$ decay}
\bigskip\bigskip\Black
\centerline{\large\bf Paolo Ciafaloni} \vspace{0.3cm}
\centerline{\em IFAE - Grup de F\'{\i}sica Te\`orica, Edifici Cn,
Universitat  Aut\`onoma}
\centerline{\em de Barcelona,
08193 Bellaterra, Espa\~{n}a and INFN -- Frascati -- Italy,}
\bigskip
\centerline{\large\bf Andrea Romanino {\rm and} Alessandro Strumia} \vspace{0.3cm}
\centerline{\em Dipartimento di Fisica, Universit\`a di Pisa and}
\centerline{\em INFN, sezione di Pisa,  I-56126 Pisa, Italia}\vspace{0.3cm}
\bigskip\bigskip\Blue
\centerline{\large\bf Abstract}
\begin{quote}\large\indent
The charged-Higgs-mediated
contribution to the Wilson coefficient
of the $b\to s \gamma$ magnetic penguin
is expected to be one of the more promising candidates for a supersymmetric
effect in $B$ physics, probably the only one in gauge-mediated models.
We compute the two-loop QCD correction to it.
With na\"{\i}ve dimensional regularization and $\overline{\hbox{MS}}$ subtraction,
for reasonable
values of the charged Higgs mass and for $\bar{\mu}=m_t$,
we find a $(10\div20)\%$
reduction of the corresponding one-loop effect.
\end{quote}\Black

\section{Introduction}
In supersymmetric models with sparticle masses
mediated by supergravity~\cite{SuGraSoft},
physics at the unification scale
leaves its imprint in the supersymmetry breaking terms.
As a consequence we expect that the sfermion mass matrices contain new
sources of flavour and CP violation,
either due to unification physics~\cite{FVGUT} or
related to the generation of the flavour structure itself~\cite{FVfla}.
In this scenario it is quite possible that the effects
due to virtual sparticle exchanges
(mostly $\mu\to e\gamma$, $\mu\to e$ conversion in atoms,
electric dipoles of the electron and of the neutron,
CP-violation in the $K$-system and $B$-systems),
will be discovered even before than the sparticles themselves.

Alternatively, in supersymmetric models, like Gauge-Mediated (GM) models,
where the supersymmetry-breaking
soft terms are generated at a lower scale 
where non-MSSM physics has decoupled, we
expect that the only flavour violation present at low energy
be the supersymmetrized extension of the standard CKM matrix.
In this scenario the new supersymmetric
flavour violating interactions
(mainly the ones at charged Higgs and higgsino vertices,
present in any realistic supersymmetric extension of the SM)
are essentially unrelated to supersymmetry breaking
and their flavour structure is controlled by the same CKM matrix.
Consequently they do not introduce any new CP-violating phase nor
they affect leptons,
so that supersymmetry gives
contributions only to `standard' flavour and/or CP violating effects.
These effects can nevertheless lead to detectable deviations from the expectations
of the standard model in a few crucial observables in flavour physics,
mainly in the $b\to s\gamma$ and $b\to s\ell^+\ell^-$ decays~\cite{C7H,FVMSSM}.
Supersymmetric effects in other processes are less interesting,
because the SM background is larger and/or plagued by larger QCD uncertainties. 

\smallskip

From the point of view of the effective theory below the Fermi scale,
in this scenario
supersymmetry is expected to give a detectable correction to the Wilson coefficient,
$C_7$, of the $b\to s \gamma$ magnetic penguin
operator, usually named $\Op{7}$.
This coefficient can be extracted from ${\rm B.R.}(b\to s\gamma)$
with a $\sim10\%$ theoretical uncertainty~\cite{NLO7,QCDbsg}.
Alternatively it can be deduced from
the spectrum of $b\to s\ell^+\ell^-$ decays away from resonances.
These decays have a lower branching ratio (around $6\cdot 10^{-6}$ for
$\ell = e,\mu$ and smaller for $\ell=\tau$) than $b\to s\gamma$.
Consequently, for a precise determination of $C_7(\mub\approx m_t)$
one needs to wait for sufficient statistics.
However, with data from $5\cdot 10^8$ $B\bar{B}$ pairs, the $1\sigma$ error on $C_7$
will again be slightly larger than $\pm 0.01$~\cite{Hew}.

In the SM the perturbative QCD uncertainties in
${\rm B.R.}(b\to s\gamma)$ have been reduced
from $\sim30\%$ to $\sim5\%$
computing the full Next-to-Leading-Order (NLO) QCD corrections.
This computation can be divided in three steps:
(i) at some scale $\mub$ close to the electroweak scale,
the couplings of the effective theory
(containing only the light degrees of freedom)
are determined up to $\Ord(\alpha_3)$~\cite{Adel,GH}
in such a way that the full and effective theories describe
equivalent physics;
(ii) the effective theory is evolved via RGE techniques at NLO
from the electroweak scale down to the $B$ scale~\cite{NLO7};
(iii) at the $B$ scale
the matrix elements for the $b\to s\gamma$ process are computed
with $\Ord(\alpha_3)$ precision~\cite{matching}.
With $\overline{\rm MS}$ subtraction, and in
the `na\"{\i}ve dimensional regularization' (NDR) scheme commonly employed,
these three parts are of comparable numerical importance~\cite{GH}.

In the scenario where supersymmetry 
affects the low energy theory only giving extra contributions
to the same Wilson coefficients that describe the SM physics,
`only' the step (i) is missing for a complete computation of
supersymmetric corrections at NLO precision.
We do this computation in the case of the charged Higgs mediated correction to $C_7$.
In the next section we explain why this correction is of particular interest.
In section~3 we collect the ingredients for the computation.
The computation in the full theory is presented in section~4,
and the one in the effective theory in section~5.
The final result is given in section~6.
For a reasonable Higgs mass, $m_H\sim 2 m_t$, we find that in the NDR scheme
the one-loop charged-Higgs contribution, evaluated at $\mub=m_t$, is
reduced by two-loop QCD corrections by $15\%$.

\section{Relevant supersymmetric effects}
Once that $C_7$ will be measured with sufficient precision,
the accurate SM computations could say that some new physics is required.
Needless to say, however, as always in the case of radiative
corrections effects, the discovery of non SM effect would not allow an
immediate identification of its physical origin.
The supersymmetric effects, in particular, depend on
many unknown parameters\footnote{We use standard notations
for the supersymmetric parameters; in particular $m_H$ is the
charged Higgs mass, $\tan\beta$ is the ratio between the
vacuum expectation values of the two Higgs fields,
and $\mu$ is the `$\mu$-parameter', not to be confused
with the $\overline{\rm MS}$ scale $\mub$.}:
the charged Higgs contribution depend on $m_H$ and $\tan\beta$;
the chargino contribution (at one loop)
depend on the $\mu$-parameter, on the weak gaugino masses,
on the trilinear term of stop squark, $A_t$, and on the up-type squark masses.
So, it does not seem useful to replace an approximate function
of many unknown parameters with a more precise function
(the NLO chargino contribution depends also
on the masses of various supersymmetric particles in a non-decoupling way,
if the couplings $\lambda_t,\lambda_b,g_2$ at chargino vertices
are expressed in terms of measured SM parameters).

To discuss this point and to assess the relative importance between the different
contributions we need to look in more detail at the one-loop predictions.
In the simplifying limit
$m_H\gg m_t$,  $m_{\tilde{c}}\gg m_{\tilde{t}}\gg \mu \gg M_W$
the relevant contributions from Standard Model (SM),
charged Higgs ($H$) and charginos ($\chi$) to the
Wilson coefficients $C_7$ of the $b\to s \gamma$ magnetic penguin, and
$C_8$ of the $b\to s g$ chromo-magnetic penguin are
\begin{equation}\label{eq:C7}
\begin{array}{ll}
\displaystyle
C_7^{\rm SM} \approx -0.2  &\displaystyle
C_8^{\rm SM} \approx -0.1  \\[3mm]
\displaystyle
C_7^{H} \approx -\frac{m_t^2}{2M_H^2}
\left(\frac{7}{36}\frac{1}{\tan^2\beta} +
 \frac{2}{3}\ln\frac{m_H^2}{m_t^2}-\frac{1}{2}  \right) &\displaystyle
C_8^{H} \approx -\frac{m_t^2}{2M_H^2}
\left(\frac{1}{6}\frac{1}{\tan^2\beta} + \ln\frac{m_H^2}{m_t^2}-\frac{3}{2}\right) \\[2mm]
\displaystyle
C_7^\chi \approx\frac{m_t^2}{2m_{\tilde{t}}^2}\frac{1}{\sin^2\beta}
\left(\frac{2}{9}+\frac{\mu (A_t\tan\beta +\mu)}{m_{\tilde{t}}^2}
(\ln\frac{m_{\tilde{t}}^2}{\mu^2}-\frac{13}{6})\right) &\displaystyle
C_8^\chi \approx\frac{m_t^2}{2m_{\tilde{t}}^2}\frac{1}{\sin^2\beta}
\left(\frac{1}{12}-\frac{1}{2}\frac{\mu (A_t\tan\beta +\mu)}{m_{\tilde{t}}^2}\right)\\[2mm]
\end{array}
\end{equation}
We see that the charged Higgs gives an important correction,
with the same sign of the SM contribution.
For example, even a heavy charged Higgs with mass $m_H=700\GeV$ gives a $10\%$
enhancement of ${\rm B.R.}(b\to s \gamma)$ over the SM prediction.
The chargino contribution is relevant if a stop state is
relatively light and the other up-squarks are heavier.
A possible gluino/squark contribution is at most few $\%$ than
the SM result, even in presence
of new CKM-like mixing matrices with elements of the order of
the CKM-ones~\cite{FVGUT,FVMSSM}.
A possible neutralino/squark contribution is even more negligible.

Moreover, the scenario where $b\to s\gamma$ seems the most interesting
candidate for a detectable supersymmetric effect can naturally be realized
in gauge-mediation models~\cite{GaugeSoft}.
In these quite predictive models
{\em the chargino-up-squark loops do not give
large contributions}, unless $\tan\beta$ is large~\cite{Dutta}.
This happens because in a typical gauge mediated spectrum the
squarks are rather heavy (their mass terms are mediated by strong gauge interactions)
and `rather' degenerate among them (gauge interactions are generation universal),
so that chargino/squark contributions are small and GIM-suppressed.
Consequently, in gauge-mediation scenarios,
the most (and, probably, only) interesting supersymmetric
effect in $B$ physics
is given by a charged-Higgs/top correction to the $b\to s\gamma$ magnetic penguin.
This extra contribution depends only on $m_H$ and $\tan\beta$, and more precisely:
\begin{itemize}

\item The dependence on $\tan\beta$ of the charged-Higgs-mediated
magnetic penguin is very weak.
As clear from\eq{C7}, the term suppressed by $1/\tan^2\beta$ is very small
already for $\tan\beta\circa{>}2$.
We remember that smaller values of $\tan\beta$, close to one,
are unnatural in the successful scenario
of radiative breaking of the electroweak symmetry~\cite{REB},
and make difficult to accommodate the large value of the top mass.

\item The charged Higgs mass $m_H$ is not predicted by gauge-mediation scenarios;
however the mass parameters of the Higgs potential determine the electroweak scale,
so that (unless $\tan\beta$ is large), barring
unnatural accidental cancellations, the charged Higgs should not
be much heavier than the $Z$ boson.

\end{itemize}
So, the charged-Higgs mediated
contribution depends almost uniquely on the charged Higgs mass
and has defined sign.
For these reasons we think that it is interesting to
have a more accurate determination of the term not suppressed
by $1/\tan^2\beta$, that arise from graphs that contain one vertex
$\lambda_b~\bar{b}_Rt_L\,H_-$.

\smallskip

More generally this computation also applies to
two-Higgs doublet models, where the charged Higgs mediated contribution
can be the only correction to the SM prediction.

\section{Preparation for the computation}
To be more precise, and to conform to the standard notations,
we recall the effective Hamiltonian for the $b\to s\gamma$ decay
\begin{equation}
{\cal H}_{\rm eff} = -\frac{g_2^2}{2 M_W^2}V_{tb} V_{ts}^*\sum_{i=1}^8
C_i \frac{{\cal O}_i}{4}.
\end{equation}
We have approximated $V_{cb} V_{cs}^* \approx -V_{tb} V_{ts}^*$ and,
in the limit $m_s\to 0$,
\begin{equation}
\begin{array}{ll}\displaystyle
{\cal O}_1 \equiv 4(\bar{s}^i\gamma_\mu \pL c^j)(\bar{c}^j\gamma_\mu \pL b^i)  \qquad& 
{\cal O}_2 \equiv 4(\bar{s}^i\gamma_\mu \pL c^i)(\bar{c}^j\gamma_\mu \pL b^j) \\
\displaystyle
{\cal O}_3 \equiv 4(\bar{s}^i\gamma_\mu \pL b^i)\sum(\bar{q}^j\gamma_\mu \pL q^j) & 
\displaystyle
{\cal O}_4 \equiv 4(\bar{s}^i\gamma_\mu \pL b^j)\sum(\bar{q}^j\gamma_\mu \pL q^i) \\
\displaystyle
{\cal O}_5 \equiv 4(\bar{s}^i\gamma_\mu \pL b^i)\sum(\bar{q}^j\gamma_\mu \pR q^j) & 
\displaystyle
{\cal O}_6 \equiv 4(\bar{s}^i\gamma_\mu \pL b^j)\sum(\bar{q}^j\gamma_\mu \pR q^i) \\
\displaystyle
{\cal O}_7 \equiv 4\frac{e}{(4\pi)^2}m_b(\bar{s}\gamma_{\mu\nu}  \pR b)F_{\mu\nu}&
\displaystyle
{\cal O}_8 \equiv 4\frac{g_3}{(4\pi)^2}m_b(\bar{s}\gamma_{\mu\nu}T^a  \pR b)G^a_{\mu\nu}
\end{array}
\end{equation}
where $i,j$ are colour indexes, and the sum is over all light quarks $q=\{b,c,s,u,d\}$.
The relevant Wilson coefficients at the electroweak scale,
in LO approximation, are
\begin{equation}\label{eq:Cs}
\begin{array}{ll}
 \displaystyle
C_2^{\rm SM}=1 & C_2^H=0\\
\displaystyle
C_7^{\rm SM} = \frac{3}{2}[q_u P_{FE}+q_W P_{BE}]& \displaystyle
C_7^H = \frac{1}{2}[(q_u P_{FI}+q_H P_{BI})+
\frac{(q_u P_{FE}+q_H P_{BE})}{\tan^2\beta}]\\
\displaystyle
C_8^{\rm SM} = \frac{3}{2} P_{FE}& \displaystyle
C_8^H=\frac{1}{2} [ P_{FI}+\frac{P_{FE}}{\tan^2\beta}]
\end{array}
\end{equation}
At leading order the magnetic penguin coefficients $C_7$ and $C_8$
are given by the value of the corresponding penguin diagrams,
without corrections from the matching procedure.
The magnetic penguin loop-functions $P_{F/B,I/E}(r)$,
that we will also
employ to write the divergent part of our two-loop Feynman graphs,
are listed in the appendix.
The indices $F$ or $B$ indicate if the external photon (or gluon)
is attached to a $F$ermion or a $B$oson.
The indices $E$ or $I$  indicate if the helicity flip factor, $m_b$, comes from
the $E$xternal $b$ leg, or from an $I$nternal vertex of the graph.
The parameter $r\equiv m_B^2/m_F^2$ is the ratio between the
(squared) masses of the boson $B$
and of the fermion $F$ in the loop ($r=M_W^2/m_t^2$ in the SM case
and $r=m_H^2/m_t^2$ in the charged Higgs case).

\medskip

We will {\em not\/} compute the NLO corrections to the $b\to s g$
chromo-magnetic penguin operator $\Op{8}$.
This would be formally inconsistent in an expansion where no
small parameter is present in the QCD RGE corrections
so that the RGE mixing between $\Op{7}$ and $\Op{8}$ is considered of $\Ord(1)$.
However the relevant RGE-loop factor is of order
$\ell\equiv (\alpha_3/4\pi) \ln M_W^2/m_b^2 \approx 0.1$, smaller than 1.
More in detail, at leading order, the coefficient $C_7$ at the $B$ scale
is obtained in terms of the coefficients at the electroweak scale as
$$C_7(m_b)\approx U_{72}+U_{77}C_7+U_{78}C_8\approx -0.155+
0.7 (-0.2+C_7^{\rm SUSY})+0.085(-0.1+C_8^{\rm SUSY})\approx
-0.30+0.7 C_7^{\rm SUSY}.$$
Here $U_{ij}$ is the evolution matrix from the weak scale to the $B$ scale
for the Wilson coefficients $C_i$.
This confirms that it is not necessary to compute $C_8$ at NLO.

More generally, the RGE-loop factors of order one, that surely
need to be exactly resummed via the RGE techniques,
are the ones that determine the running of the strong coupling constant.
This happens because the QCD $\beta$ function receives contributions
from {\em all} (`active') quarks,
while only some specific flavours contribute, \hbox{i.e.},
to the mixing between the $b\to s$ penguins $\Op{7}$ and $\Op{8}$.

Consequently, when studying the decay $b\to s\ell^+\ell^-$,
it is not inconsistent to add the correction to $C_7$
we are going to compute (together with the corresponding low energy part),
even if not required by a `formal' expansion at NLO~\cite{Burassone}.

This correction has already been computed in~\cite{Anlauf} with
effective theory techniques, but only in the
limit $m_H\gg m_t$ where the effect is negligible.
The complete NLO computation has instead been done in the case
of charged Higgs corrections to the $B_0\bar{B}_0$ mixing~\cite{MixNLO}.

\medskip

We can now pass to the NLO computation of the part of the
charged-Higgs mediated contribution
to $C_7$ not suppressed by $1/\tan^2\beta$.
This computation requires to match the NLO (two-loop) $b\to s\gamma$ amplitude
in the full theory with the one of the effective theory.
We choose to employ the equations of motion $(i\ds b=m_b b)$
so that $\Op{7}$ is the only relevant operator.
In this way we are forced to match {\em on-shell} $b\to s\gamma$ amplitudes,
plagued by infrared divergences that must be properly treated.

In both versions of the theory we choose the Feynman gauge for the gluon propagator.
Since we never need to define traces
like $\hbox{Tr}\,\gamma_5\gamma_\mu\cdots \gamma_\rho$,
nor the completely antisymmetric tensor,
we can employ na\"{\i}ve dimensional regularization
(i.e.\ anticommuting $\gamma_5$)
with $\overline{\hbox{MS}}$ renormalization scale $\mub$.
This regularization is more convenient than the supersymmetry-preserving
`dimensional reduction'.
Infact the contribution we want to compute
does not depend on supersymmetric parameters (like gaugino and sfermion masses,\ldots)
that can be related among them making assumptions about physics at high energy,
giving rise to predictions usually computed employing dimensional reduction.

For simplicity in the following we
denote with $C_7$, $C_8$,\ldots, only
their charged-Higgs mediated part that we want to compute.

\section{$b\to s\gamma$ in the full theory}
As said we are interested in diagrams that contain one $\lambda_b~\bar{b}_Rt_L\,H_-$
vertex.
We have to compute the 16 two-loop diagrams shown in figure~1.

We write the various contributions $\Gamma$
to the two-loop $b\to s\gamma$ effective Hamiltonian in the full theory as
\begin{equation}
{\cal H}_{b\to s\gamma} = -\frac{g_2^2}{2 M_W^2}\cdot V_{tb} V_{ts}^*
\cdot (\hat{C}_{70}+\frac{\alpha_3}{4\pi}c_3\hat{\M}) \frac{\Op{7}}{4},\qquad
\hat{\M}= \sum_\Gamma \hat{\M}_\Gamma
\end{equation}
where $c_3 = 4/3$ is the quadratic Casimir for the fundamental
representation of $\SU(3)_c$,
and the overall factor has been defined in such a way that the
$c_3\hat{\M}$ is normalized as $C_{71}$,
the NLO term of the Wilson coefficient $C_7$
of the operator $\Op{7}$ in the effective theory:
$$C_7 = C_{70}+ \frac{\alpha_3}{4\pi}C_{71}+\Ord(\alpha_3^2).$$
The LO coefficient of $\Op{7}$ in
the full theory, $\hat{C}_{70}$, differs from the corresponding one in the
effective theory, $C_{70}$, by terms that vanish as $d\to 4$,
as described in eq.\eq{hatC} in the appendix.
Since the parameters of the theory receive divergent corrections,
these terms will give a finite contribution to $C_{71}$.

\begin{figure}[t]\setlength{\unitlength}{1cm}
\begin{center}
\begin{picture}(13,6.3)
\ifMac
{\put(-0.5,0){\special{picture feyn}}}
{\put(-1,-2.3){\includegraphics{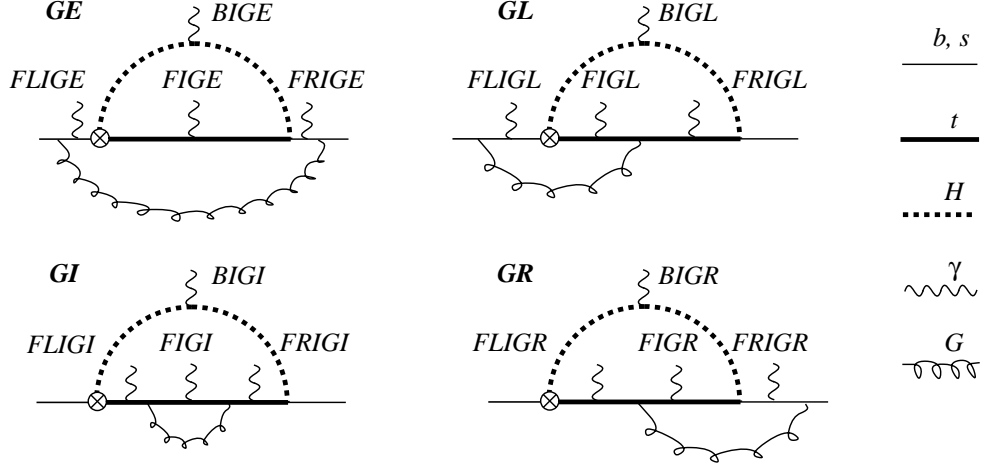}}}
\end{picture}
\caption[SP]{\em The 16 two-loop Feynman graphs
(we show on a single diagrams all possible attachment of the photon
and write near to it the name we give to the diagram) that we have to compute.
The symbol $\otimes$ denotes the vertex $\bar{b}_R t_L H_-$.
\label{fig:feyn}}
\end{center}\end{figure}

\subsection{Two loop diagrams}
A given Feynman diagram $\Gamma$ gives a contribution $\hat{\M}_\Gamma=\int F_\Gamma$
(where $\int$ denotes the standard two-loop integration over internal loop momenta)
that depends on the heavy masses,
$M=\{m_t,m_H\}$ and on the light masses and momenta, $m=\{m_b,m_s;p_b,p_s\}$,
of the $b$ and $s$ quarks.
The full expression $\hat{\M}_\Gamma(M,m)$ contains negligible
terms suppressed by powers of $m/M$.
The natural trick that allows to avoid computing all the unnecessary terms,
without loosing terms like $\ln m/M$,
consists in adding and subtracting to the loop-integrand $F_\Gamma(M,m)$
some appropriately chosen {\em simpler} term $\Delta F_\Gamma(M,m)$ that has
the same low energy structure of the full $F_\Gamma(M,m)$.
The result without irrelevant powers of $m/M$ is given by the simpler integral
$\int[F_\Gamma(M,0)-\Delta F_\Gamma(M,0)]+\Delta F_\Gamma(M,m)$.
In practice {\em simpler} means that $\Delta F_\Gamma$ can be chosen
as a product of two one-loop integrals, instead of a two-loop integral,
so that, in dimensional regularization,
$\int\Delta F(M,0)=0$.

It is convenient to employ a standard technique~\cite{HME}, usually named
{\em `Heavy Mass Expansion'\/} (HME),
that generalises and systemizes this procedure.
It is possible to show that,
in dimensional regularization, the expansion of a Feynman diagram $\Gamma$,
that depends on `heavy' masses and momenta $M$
and on `light' masses and momenta $m$,
in the limit $M\gg m$ can be done at the level of integrands as
\begin{equation}
F_{\Gamma}(M,m) \simeq
\sum_{\gamma\subseteq\Gamma} F_{\Gamma / \gamma}(m)~{\cal T}_{q_\gamma,m_\gamma} 
F_{\gamma}(q_\gamma,m_\gamma,M)\qquad\hbox{as $M\gg m$}
\end{equation}
where ${\cal T}_x$ indicates Taylor expansion in the variables $x$
up to the desired order.
The sum is performed over all the ``asymptotically irreducible''
subgraphs $\gamma$ of $\Gamma$, i.e.\ those which 
satisfy the following two conditions:
\begin{itemize}
\item[1.] $\gamma$ contains all the lines with heavy masses;
\item[2.] $\gamma$ consists of connectivity components that are
one-particle-irreducible with respect to lines with small masses.
\end{itemize}
Finally $q_\gamma$ denotes the set of momenta external to the subgraph $\gamma$,
$F_\gamma$ is its Feynman integrand,
and $F_{ \Gamma / \gamma}$ is the Feynman integrand of the reduced graph
$\Gamma/\gamma$.
We refer the reader to~\cite{GH,HME} for explanations and examples.
Here we discuss how this technique works in our case,
and its relation with the natural trick previously discussed.
\begin{itemize}
\item[-]
The case $\gamma=\Gamma$ is always present and gives a `na\"{\i}ve' Taylor
expansion in $m$.
This contribution corresponds to the term $F_\Gamma(M,0)$
of our previous example.

\item[-] There is one non-vanishing contribution from the
`non na\"{\i}ve' part of the HME expansion from
each one of our graphs in fig.~1
(except the {\em ``GI''\/} diagrams with an `internal' gluon loop).
The ``asymptotically irreducible'' subgraph $\gamma$
coincides with the Higgs/top loop that gives the LO result.
This term of the HME expansion
can be seen, at a diagrammatic level,
as the contribution of the full diagram with the heavy propagators contracted to a point,
and corresponds, in our previous example, to the term $\Delta F_\Gamma(M,m)$.
When computing it we neglect powers of $\rho\equiv m_s^2/m_b^2$.

\item[-] Terms analogous to $-\Delta F_\Gamma(M,0)$ are not included
in this form of the HME, since they vanish in dimensional regularization.
For this reason some fake {\em infrared\/} (IR) divergences can (and will)
appear in the `na\"\i{}ve part' where $m=0$.
These divergences are cancelled diagram by diagram by fake
{\em ultraviolet\/} (UV) divergences
that appear in the non na\"\i{}ve part of the HME expansion.
\end{itemize}
The `na\"\i{}ve part' of the expansion
also contains the usual UV divergences that cancel upon renormalization.
Similarly, the `non na\"\i{}ve part' also contains IR divergences,
cancelled in the final physical result
after phase-space integration and inclusion of QCD bremsstrahlung $b\to s\gamma g$.

In order to check in detail all these cancellations, we
have separated IR from UV divergences.
It has been possible to do it in dimensional regularization,
since our graphs have only single poles $1/\eps$.
In the following we will denote as $1/\epsUV$ an ultraviolet pole,
and as $1/\epsIR$ a pole of infrared origin.

In our case, since we have already a $m_b$ factor from the vertex,
the Taylor expansions have to be performed up to first order
(up to second order in the `non-na\"{\i}ve part'
of the diagrams `$FLIGE$' and `$FRIGE$' with four light propagators).
The `na\"{\i}ve' parts of diagrams that differ by the exchange
$b\rightleftharpoons s$ (like the diagrams `$FLIGE$' and `$FRIGE$')
are equal because the
Dirac equation, that sees the difference between the $b$ and $s$ masses,
is never used.

\medskip

We have written a {\tt Mathematica} code that simplifies the spinor algebra,
expands the `light' factors in the loop integrands in Taylor expansion
up to the appropriate order,
reducing the full expression
to a sum of one and two-loop scalar integrals
that have been separately computed.
This technique is less efficient than the alternative one
based on Feynman parameters; however it is easy to teach it to a computer.
The non-na\"{\i}ve part of the HME expansion
is computed in the same way, performing the Taylor expansion also in
one (appropriately chosen) loop momentum.
In this way the whole computation
is done in few minutes by a normal `personal' computer.
We show the separate result of the two parts of the
expansion ($\hat{\M}_{\rm 2~loop}$ from the `na\"{\i}ve' (high-energy) part,
and $\hat{\M}_{\rm HME}$ from the `non na\"{\i}ve'
(low energy) term of the HME expansion)
in eq.s~(\ref{sys:hatM}).
We have listed in table~1 all the divergences present in the single graphs.

\begin{table}
$$\begin{array}{|lc|cc|cc|}\hline
\multicolumn{2}{|c|}{\hbox{graph}\hfill\hbox{coefficient}}&
\multicolumn{2}{|c|}{\hbox{na\"{\i}ve part of the HME expansion}}&
\multicolumn{2}{|c|}{\hbox{non na\"{\i}ve part of the HME expansion}}\\
\multicolumn{2}{|c|}{\hbox{and charge}\downarrow\hfill\hbox{of}\to}&
1/\epsUV & 1/\epsIR & 1/\epsUV
& \ln( m_s^2/m_b^2)/\epsIR \\ \hline 
BIGE &q_H& 0 &-P_{BI} &+P_{BI}&-\frac{1}{2}P_{BI}\\ 
BIGI &q_H&-2P_{BI}-3r P'_{BI}& 0 &0&0\\
BIGL,BIGR &q_H&2 P_{BI} & \frac{1}{2}P_{BI}&-\frac{1}{2}P_{BI}&0\\  \hline
FIGI&q_u& +\frac{1}{2}P_{FI}&0&0&0\\
FLIGI,FRIGI&q_u&-\frac{5}{4} P_{FI}-\frac{3}{2}rP'_{FI}&0&0&0\\
FLIGR,FRIGL&q_u&2P_{FI}&+\frac{1}{4} P_{FI}&-\frac{1}{4}P_{FI}&0\\
FIGL,FIGR&q_u&0&-\frac{1}{4}P_{FI}+\frac{1}{2}P_{BI}&+\frac{1}{4}P_{FI}-\frac{1}{2}P_{BI}&0\\
FIGE&q_u&0&-P_{BI}&P_{BI}&-\frac{1}{2}P_{FI}\\ \hline
FRIGR,FLIGL&q_d&0&P_{FI}-\frac{1}{2}P_{BI}&-P_{FI}+\frac{1}{2}P_{BI}&0\\
FLIGE,FRIGE&q_d&0&\frac{1}{2}P_{BI}&-\frac{1}{2}P_{BI}&0\\ \hline\hline\Blue
\hbox{\bf all graphs $\propto$}&q_H&2P_{BI}-3rP'_{BI} & 0 &0&-\frac{1}{2} P_{BI}\\
\hbox{\bf all graphs $\propto$}&q_u&2P_{FI}-3rP'_{FI} & 0 &0&-\frac{1}{2} P_{FI}\\
\hbox{\bf all graphs $\propto$}&q_d&0 & 2 P_{FI} &-2P_{FI}&0\\ \hline
\end{array}$$\Black
\caption{\em Coefficients of the divergent parts of the single graphs.
The loop functions $P_{BI}$ and $P_{FI}$ are defined in the appendix;
the `names' of the graphs are defined by fig.~1.}\end{table}

\subsection{Renormalization}
It is necessary to renormalize the top mass, $m_t$, the bottom mass, $m_b$,
and to express the result in terms of
canonically normalized $b$ and $s$ quark fields.

The simplest way of renormalizing the top mass consists in
expressing the bare top mass, $m_{t0}$, in term of the renormalized `running'
mass, $m_t$,
$$m_t = m_{t0}+\delta m_t,\qquad
\frac{\delta m_t}{m_t} = c_3\frac{\alpha_3}{4\pi}\frac{3}{\epsUV}$$
($c_3\equiv 4/3$) in the one loop contribution, $\hat{C}_{70}(r)$.
The corresponding contribution $\hat{\M}_t$ to the amplitude $\hat{\M}$ is
\begin{eqnsystem}{sys:ctr}
\hat{\M}_t &=& \frac{3}{\epsUV}2r \hat{C}'_{70}\\
\riga{where $'$ denotes derivation with respect to $r=m_H^2/m_t^2$.
The pole $b$-quark mass $m_b$ is expressed
in terms of the bare $b$ mass, $m_{b0}$, as
$$m_b = m_{b0}+\delta m_b,\qquad
\frac{\delta m_b}{m_b} = c_3\frac{\alpha_3}{4\pi}(\frac{3}{\epsUV}
+3\ln\frac{\mub^2}{m_b^2}+4).$$
Since in the one-loop graphs the $m_{b0}$ factor
comes only from the vertex (we never use the Dirac equation to reduce the
operator basis), the contribution $\hat{\M}_b$ to $\hat{\M}$
from the renormalization of $m_b$ is simply given by}\\
\hat{\M}_b &=& -\hat{C}_{70}(\frac{3}{\varepsilon}+3\ln\frac{\bar{\mu}^2}{m_b^2}+4).\\
\riga{The counterterms for {\em on-shell} renormalization of the wave-function 
of the $b$ and $s$ quarks are
$$Z_{q}= 1-c_3\frac{\alpha_3}{4\pi}
(\frac{1}{\epsUV}+\frac{2}{\epsIR}+3\ln\frac{\bar{\mu}^2}{m_q^2}+4),\qquad
q=\{s,b\}.$$
Expressing the one-loop result in terms of canonically
normalized $b$ and $s$ quark fields, we get the following correction
$\hat{\M}_Z$ to $\hat{\M}$}\\
\hat{\M}_Z &=&- \hat{C}_{70}
(\frac{1}{\varepsilon}+\frac{2}{\epsIR}+3\ln\frac{\bar{\mu}^2}{m_b m_s}+4).
\end{eqnsystem}
Finally, top/Higgs loops generate non diagonal
kinetic and mass terms between the $b$ and $s$ quarks
that is necessary to eliminate via appropriate flavour rotations of the quark fields.
This gives rise to a contribution to $\hat{\M}$ when a magnetic $bb\gamma$ dipole operator
is rotated into a $bs\gamma$ dipole.
However, it is more convenient to take into account this particular correction
adding the (one$+$one)-loop diagram of figure~2
(the similar graph with $b\rightleftharpoons s$ is suppressed by powers of $m_s/m_b$)
to the list of the two-loop diagrams.

We denote the renormalization contributions as
$\hat{\M}_{\rm ctr} = \hat{\M}_t+\hat{\M}_b+\hat{\M}_Z$.

\begin{figure}[t]\setlength{\unitlength}{1cm}
\begin{center}
\begin{picture}(4.5,1.3)
\ifMac
{\put(-0.5,0){\special{picture feyn11}}}
{\put(-6,-4.8){\includegraphics{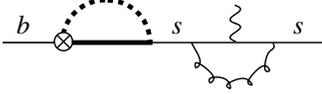}}}
\end{picture}\hspace{0.7cm}\raisebox{1.8cm}{
\parbox[t]{7.5cm}{\caption{\em The Feynman graph that
accounts for the renormalization of
the mass and kinetic terms that mix the $b$ and $s$ quarks.\label{fig:feyn11}} }}
\end{center}\end{figure}

\section{$b \to s \gamma$ in the effective theory}
Since all the supersymmetric contributions (if $R$-parity is conserved)
arise first at one loop level, their mixing structure is simpler
than the one of the SM contributions.
In particular all the LO supersymmetric contributions to the Wilson coefficients $C_i$
are given by the corresponding penguin diagrams
without extra matching contributions.
We list here all the various contributions to the $b \to s \gamma$ decay
amplitude in the effective theory relevant for our computation
(we are neglecting terms suppressed by $\tan^{-2}\beta$).
We denote by $\M$ the value of the on-shell $b\to s \gamma$ amplitude
in the effective theory,
normalized in the same way as $\hat{\M}$ (the corresponding amplitude
in the full theory).
The non zero contributions to $\M$ are, in our case:
\begin{itemize}
\item The contribution $\M_7$ given by the
matrix element of $\Op{7}$ itself at order $\alpha_3$
\begin{equation}\label{eq:rhat7}
\M_7 = \frac{1}{c_3}  C_{71} -C_{70} \ln\rho \left[
\frac{1}{\epsIR}+\ln\frac{\mub^2}{m_b^2} +2-
\frac{1}{2}  \ln\rho \right].
\end{equation}
The first term is the tree-level matrix element
of the term we ultimately
wish to extract.
The other terms are obtained by the one-loop QCD correction to
the leading order magnetic penguin $\Op{7}$.

\item The order $\alpha_3$ matrix element of the chromo-magnetic
penguin operator $\Op{8}$
\begin{equation}
\M_8 = q_d C_{80}
\left[-\frac{4}{\epsUV}-4\ln\frac{\mub^2}{m_b^2} - 11 + \frac{2\pi^2}{3}
- 2i\pi\right].
\end{equation}

\item As the operators mix under renormalization, we have to consider
counterterm contributions induced by operators of the form
$C_i \, \delta Z_{ij} \bAk{s\gamma}{\Op{j}}{b}$.
Using the known renormalization constants~\cite{count},
the non-vanishing contributions to the $b\to s\gamma$ amplitude $\M$ are
$$
\M_{77} =  \frac{4}{\epsUV} C_{70},\qquad\hbox{and}\qquad
\M_{87} = q_d\frac{4}{\epsUV} C_{80}.
$$
`Evanescent' operators do not give contributions.

\item
Like in the full theory, is of course necessary to
renormalize also the parameters and fields of the theory,
that in this case are the $b$-quark mass and the $b$ and $s$
wave-functions.
The counterterm due to the $b$-quark mass renormalization is
\begin{equation}
\label{deltahatrb}
\M_b = -C_{70} \left[\frac{3}{\epsUV}+3\ln\frac{\mub^2}{m_b^2} +4\right] 
\end{equation}
when using the pole $b$-quark mass as in the full theory, while
the renormalization of the $b$ and $s$ fields gives a contribution $\M_Z$ 
\begin{equation}\label{hatMZ2}
\M_{Z} = -C_{70} \left[\frac{1}{\epsUV} +\frac{2}{\epsIR} +
3\ln\frac{\mub^2}{m_b m_s}+4 \right].
\end{equation}
\end{itemize}
The amplitude $\M$ in the effective theory is UV finite and has
an IR divergence $-(2+\ln\rho)C_{70}/\epsIR$.

\section{Matching and final result}
We now collect all the necessary terms and illustrate the matching procedure.
The amplitude in the effective theory is
$$\M=\frac{1}{c_3}C_{71}+
q_d C_{80}\left[\frac{2\pi^2}{3}-11-2i\pi-4\ln\frac{\mub^2}{m_b^2}\right]
-C_{70}\left[(2+\ln\rho)\frac{(\mub^2/m_b^2)^\varepsilon}{\epsIR}+
8+4\ln\frac{\mub^2}{m_b^2}+\frac{1}{2}\ln\rho-\frac{1}{2}\ln^2\rho\right].$$
The amplitude $\hat{\M}$ in the full theory is the sum of the contribution
$\hat{\M}_{\rm 2~loop}$,
defined as the `na\"\i{}ve' part of the two-loop diagrams,
of the `non-na\"\i{}ve' part, $\hat{\M}_{\rm HME}$, and of the counterterms,
$\hat{\M}_{\rm ctr}$
\begin{eqnsystem}{sys:hatM}
\hat{\M}_{\rm 2~loop}&=&(-\frac{6r\hat{C}'_{70}-4\hat{C}_{70}}{\epsUV}+
4q_d \frac{\hat{C}_{80}}{\epsIR})
\left(\frac{\mub^2}{m_t^2}\right)^{\!1\eps} + f(r)\\
\hat{\M}_{\rm HME} &=&-\hat{C}_{70}\left[\frac{(\mub^2/m_b^2)^{\eps}}{\epsIR}\ln\rho-
\frac{1}{2}\ln^2\rho+2\ln\rho\right]+
q_d \hat{C}_{80}\left[\frac{2\pi^2}{3}-11-
2i\pi-4\frac{(\mub^2/m_b^2)^{\eps}}{\epsUV}\right]\\
\hat{\M}_{\rm ctr}&=&\frac{6r\hat{C}'_{70}-4\hat{C}_{70}}{\epsUV}-
\hat{C}_{70}\left[2\frac{(\mub^2/m_b^2)^\varepsilon}{\epsIR}+
8+4\ln\frac{\mub^2}{m_b^2}-\frac{3}{2}\ln\rho\right]
\end{eqnsystem}
where the two-loop function $f(r)$ is, for arbitrary electric charges
$q_d=q_H+q_u$,
\begin{eqnarray}\label{eq:f(r)}
f(r)&=& \!\phantom{+}q_H\left[
\frac{17\,r -3 - 2\,{r^2}}{(r-1)^3} + 
\frac{1 - 20r - 5r^2}{2(r-1)^4}r\ln r +2
\frac{1 - 4r - {r^2}}{(r-1)^3}\Li(1-r)  \right]+\\
&&+q_u\left[3
\frac{4r-1 + 5{r^2}}{2(r-1)^3} + 
\frac{r-1- 12{r^2}}{(r-1)^4}r\ln r + 
\frac{2r -1- 9{r^2}}{(r-1)^3}\Li(1-r)\right], \nonumber
\end{eqnarray}
We have neglected higher powers in $\eps$ and $\rho\equiv m_s^2/m_b^2$, and
we have included in $\hat{\M}_{\rm HME}$ the diagram in figure~2.
The full amplitude $\hat{\M}$ is
\begin{eqnarray*}
\hat{\M}&=&f(r)+(4C_{70}-6r C'_{70}+ 4q_d C_{80})\lnEps+
q_d \hat{C}_{80}\left[\frac{2\pi^2}{3}-11-2i\pi-4\ln\frac{\mub^2}{m_b^2}\right]+\\
&&-\hat{C}_{70}\left[(2+\ln\rho)
\frac{(\mub^2/m_b^2)^\varepsilon}{\epsIR}+
8+4\ln\frac{\mub^2}{m_b^2}+\frac{1}{2}\ln\rho-\frac{1}{2}\ln^2\rho\right]+
\Ord(\eps).
\end{eqnarray*}
The on-shell $b\to s\gamma$ amplitudes
($\hat{\M}$ in the full theory and $\M$ in the effective theory)
are infrared-divergent.
The correct matching is obtained requiring that the infrared safe decay rate
obtained including QCD bremsstrahlung $b\to s\gamma g$
be the same in both (full and effective) descriptions. This gives~\cite{GH,matching}
$$\hat{\M}+\hat{\M}_{\rm ir}=\M+\M_{\rm ir}$$
where
\begin{equation}
\hat{\M}_{\rm ir}=(2+\ln\rho)\hat{C}_{70}\frac{(\mub/m_b)^2}{\epsIR},\qquad
\M_{\rm ir}=(2+\ln\rho)C_{70}\frac{(\mub/m_b)^2}{\epsIR}.
\end{equation}
So, the final result for the NLO charged Higgs contribution is
(setting $q_H=-1$ and $q_u=2/3$)
\begin{eqnsystem}{sys:C71}
C_{71}&=&c_3\left [f(r)+(4C_{70}-6r C'_{70}+ 4q_d C_{80}) \nonumber
\ln\frac{\mub^2}{m_t^2}\right]=\frac{4}{3}\frac{ 2 - 13r + 7{r^2} }{(r-1)^3} - 
\frac{2r}{9}  \frac{7 - 64r + 33{r^2}}{(r-1)^4}\ln r + \\
&&- 
\frac{16}{9}\frac{2 - 7r + 3{r^2}}{(r-1)^3}\Li(1 - r)+\left(
\frac{2}{9} \frac{8 - 47r + 21{r^2}}{(r-1)^3} + 
\frac{4r}{9}\frac{3 + 14r - 8{r^2}}
{(r-1)^4}\ln r \right)\ln\frac{\mub^2}{m_t^2}.\\[1mm]
\riga{In the limits $m_H=m_t$ and $m_H\gg m_t$
the charged Higgs contribution becomes}\\
C_7(r\to1) &=& -\frac{7}{36} + \frac{\alpha_3}{4\pi}\left[\frac{181}{81}
-\frac{35}{27}\lnEps\right],\\
C_7(r\to\infty)&=&\frac{1}{4r}-\frac{\ln r}{3r}+\frac{\alpha_3}{4\pi}\frac{1}{9r}\left[
2( 42 + 4\pi^2 - 33\ln r + 12\ln^2 r)  +(42 - 32\ln r) \lnEps\right].
\end{eqnsystem}
For  $m_H=500\GeV$ and $\mub=m_t$ the term we have computed
gives a $-17\%$ correction to the LO value.

\section{Conclusion}
In a class of supersymmetric scenarios, i.e.\ in gauge mediated models,
the charged Higgs mediated
contribution to the $b\to s\gamma$ magnetic penguin
is expected to be the most interesting supersymmetric effect in $B$-physics.
The NLO QCD correction to its Wilson coefficient $C_7(\mub)$,
that we have computed employing
na\"{\i}ve dimensional regularization with
$\overline{\rm MS}$ subtraction,
gives a $-(10\div 20)\%$ correction to the corresponding LO result
(for $\mub=m_t$ and realistic charged Higgs masses).

\paragraph{Note added:} 
Together with our article, another one containing the same computation~\cite{CDGG}
has recently appeared.
Our results are in perfect agreement, even if the conclusions are apparently different.
In particular the authors of~\cite{CDGG} claim that,
in the usual two-Higgs doublet model,
the inclusion of NLO effects
enhances the $\hbox{B.R.}(b\to s\gamma)$ with respect to its LO value.
This is due to the already known NLO enhancement
in the SM contribution~\cite{NLO7,QCDbsg,Burassone}.
On the contrary, the new correction to $C_7(\mub\approx m_t)$
that we have computed tends to reduce the
charged-Higgs mediated correction.
As discussed in the text, we consider $C_7(\mub\approx m_t)$ a
more useful phenomenological quantity than $\hbox{B.R.}(b\to s\gamma)$.

\paragraph{Acknowledgements}
One of us (A.\ R.) wishes to thank the Physics Department of the
Technical University of Munich for the warm hospitality when this work
was started and in particular A.\ Kwiatkowski and N.\ Pott for several
illuminating discussions.

\appendix
\setcounter{equation}{0}
\renewcommand{\theequation}{\thesection.\arabic{equation}}

\section{Useful functions}
The penguin one-loop functions $P(r)$ employed in eq.s\eq{Cs} are given by
$P=\hat{P}|_{\eps\to 0}$, where
\begin{eqnarray*}
\hat{P}_{BI}&=&( 1 + \varepsilon \lnEps)
\frac{r^2-1- 2r\ln r}{2(r-1)^3}+\varepsilon
\frac{3(r^2 -1)- 2r(2 + r)\ln r + 2r\ln^2 r}{4(r-1)^3} \\%
\hat{P}_{FI}&=&(1 + \eps \lnEps)
\frac{1 - 4\,r + 3\,{r^2} - 2\,{r^2}\ln r }{2(r-1)^3}+\eps
{\frac{ 1 - 8\,r + 7\,{r^2} - 6\,{r^2}\ln r + 2r^2\ln^2 r}{4\,{{\left( r-1\right) }^3}}}\\
\hat{P}_{BE}&=&(1 + \eps \lnEps)
\frac{ 2 + 3r - 6{r^2} + {r^3} + 6r\ln r}{12(r-1)^4}+\eps
\frac{ 22 + 27r - 54{r^2} + 5{r^3} - 6r(-6 - 6\,r + {r^2})\ln r - 18r\ln^2 r}{72(r-1)^4}\\
P_{FE}&=&(1 + \eps \lnEps)
\frac{6r-1-3r^2-2r^3+6r^2\ln r}{12(r-1)^4}-\eps
\frac{5 - 54r + 27r^2 + 22r^3 - 6r^2( 9 + 2r)\ln r + 18r^2\ln^2 r}{72(r-1)^4}
\end{eqnarray*}
The charged-Higgs mediated one-loop result for the coefficients
$\hat{C}_{70}$ of the $b\to s \gamma$ magnetic penguin, and
$\hat{C}_{80}$ of the $b\to s g$ chromo-magnetic penguin are~\cite{C7H,FVMSSM}
\begin{equation}\label{eq:hatC}
\hat{C}_{70} = \frac{1}{2}(q_H \hat{P}_{BI}+q_u \hat{P}_{FI})+\Ord(\eps^2),
\qquad
\hat{C}_{80}=\frac{1}{2}\hat{P}_{FI}+\Ord(\eps^2).
\end{equation}
The bi-logarithmic function, $\Li(x)$, employed in eq.\eq{f(r)} is defined as
$$\Li(x)\equiv -\int_0^x\frac{\ln (1-\xi)}{\xi}\,d\xi.$$

\small~

\end{document}
\\
Title: Two-loop QCD corrections to charged-Higgs-mediated $b\to s\gamma$ decay
Authors: P. Ciafaloni, A. Romanino and A. Strumia
Comments: 10 pages, 2 figures.
Misprints corrected. Comparison with hep-ph/9710335 added.
Report-no: IFUP-TH 48/97
\\
The charged-Higgs-mediated contribution to the Wilson coefficient
of the $b\to s\gamma$ magnetic penguin
is expected to be one of the more promising candidates for a supersymmetric
effect in B physics, probably the only one in gauge-mediated models.
We compute the two-loop QCD correction to it.
With naive dimensional regularization and MSbar subtraction,
for reasonable values of the charged Higgs mass and for mu-bar = m_top,
we find a (10--20)
\\